\documentclass[review]{elsarticle}

\usepackage{lineno,hyperref}
\journal{Nuclear Instruments and Methods}

\begin{document}

\begin{frontmatter}

\title{Characterization of the ETEL D784UKFLB 11 inch Photomultiplier Tube}

%\author{R.Svoboda\fnref{myfootnote}}
%\address{University of California, Davis, Davis CA 95616, USA}


\author[Penn]{N.Barros}
\address[Penn]{University of Pennsylvania, Philadelphia, PA 19104, USA}

\author[Penn]{T.Kaptanoglu}

\author[Muhlenberg]{B. Kimmelman}
\address[Muhlenberg]{Muhlenberg College, Allentown, PA 18104, USA}

\author[Penn]{J.R. Klein}

\author[Davis]{E. Moore}
\address[Davis]{University of California, Davis. Davis, CA 95616, USA}

\author[Davis]{J. Nguyen}

\author[Penn]{K.Stavreva}

\author[Davis]{R. Svoboda\corref{mycorrespondingauthor}}
\cortext[mycorrespondingauthor]{Corresponding author}
\ead{rsvoboda@physics.ucdavis.edu}

\begin{abstract}
Water Cherenkov and scintillator detectors are a critical tool for neutrino physics. Their
large size, low threshold, and low operational cost make them excellent detectors for long baseline neutrino oscillations, proton decay, supernova and solar neutrinos, 
double beta decay, and ultra-high energy astrophysical neutrinos. Proposals for a new generation of large detectors rely on the availability of large format, fast, cost-effective photomultiplier tubes. The Electron Tubes Enterprises, Ltd (ETEL) D784KFLB 11 inch
Photomultiplier Tube has been developed for large neutrino detectors. We have measured the timing characteristics, relative efficiency, and magnetic field sensitivity of the first fifteen prototypes.
\end{abstract}

\begin{keyword}
%\texttt{elsarticle.cls}\sep \LaTeX\sep Elsevier \sep template
%\MSC[2010] 00-01\sep  99-00
Photomultiplier tube, neutrino detectors, optical detectors, photon
detection, single phase dark matter
\end{keyword}

\end{frontmatter}

%\linenumbers

\section{Introduction}

Progress in neutrino physics has gone hand-in-hand with the development of large, monolithic optical detectors using
water or liquid scintillator both as target and active detector medium. The IMB \cite{IMB} and Kamiokande \cite{Kamiokande} detectors 
found the first evidence for neutrino oscillations in atmospheric neutrinos, later confirmed by Super-Kamiokande \cite{SKNIM}. These two detectors
also detected the burst of neutrinos from supernova 1987A. Later, the SNO detector
 \cite{SNO} showed that electron neutrinos from the sun were changing flavor, and the KamLAND detector
\cite{KamLAND} showed that electron anti-neutrinos energy spectrum from distant nuclear reactors had spectral 
distortions consistent with neutrino oscillations. Taken together, it was then apparent that that neutrino flavor
oscillations were also happening to the solar neutrinos. More recently, moderately sized liquid scintillator 
optical detectors were used to measure the neutrino mixing angle $\theta_{13}$ \cite{DCPRL1,RENO,DayaBay}, and in the new
field of neutrino astrophysics  the BOREXINO deetctor \cite{BOREXINO} has measured low energy solar neutrinos from the sun with unprecedented precision
and the ICE CUBE optical module array \cite{ICECUBE} has discovered the existence of high energy astrophysical neutrinos.

A new generation of experiments is being proposed to follow up on these discoveries. Examples include SNO+ \cite{SNOPLWP}, Hyper-Kamiokande \cite{HKWP}, 
Theia \cite{ASDCWP}, CHIPS \cite{CHIPS}, and ANNIE \cite{ANNIELOI}. The availability of large format, fast photomultiplier tubes (PMTs)  is critical for
these experiments. This report describes the characterization of the timing, efficiency, and magnetic sensitivity of a new 11-inch PMT (designated
model D784UKFL) being developed by Electron Tube Enterprises Ltd (ETEL).

\section{The D784UKFL PMT}
\begin{center}
\begin{figure}[h]
\includegraphics[width=0.9\textwidth]{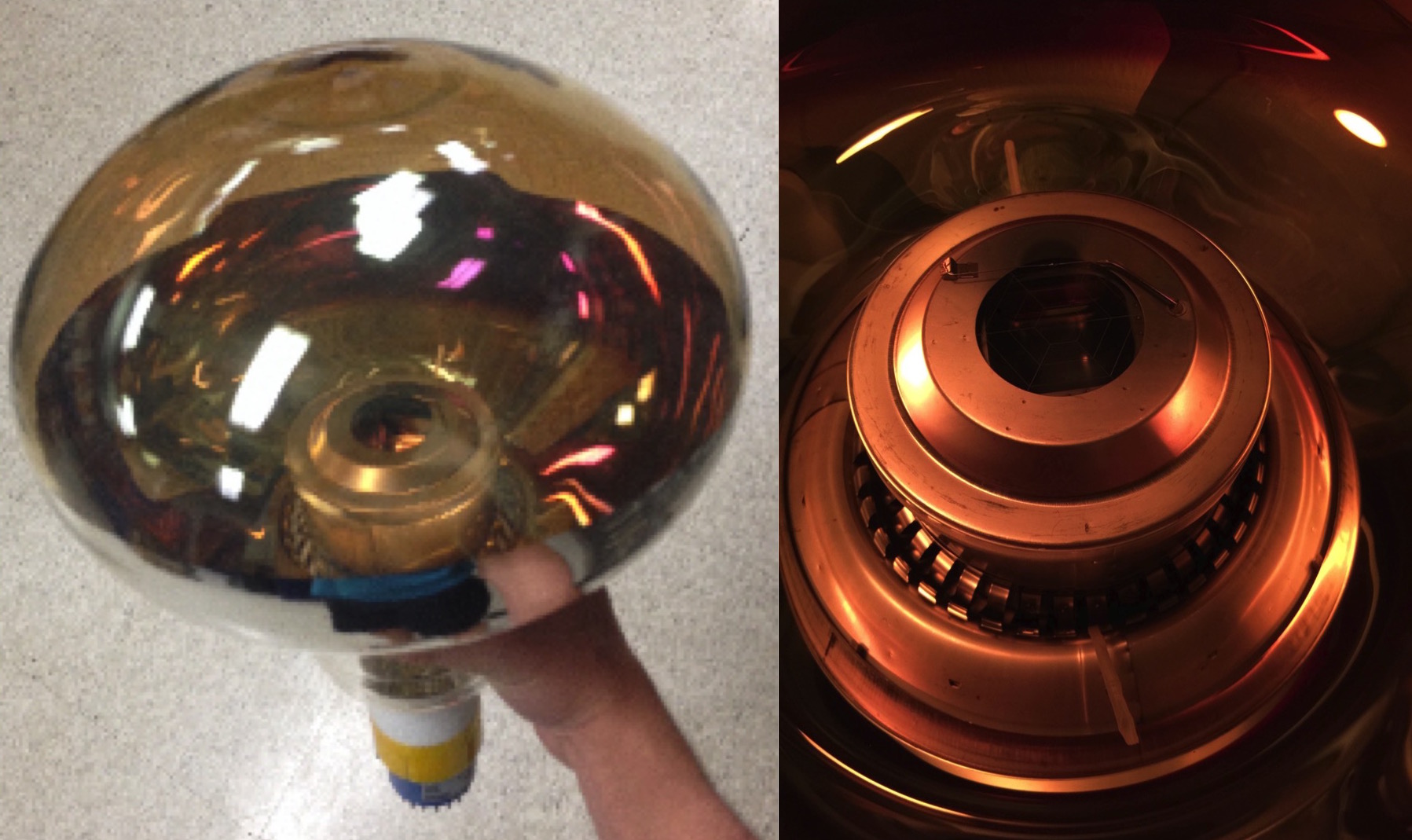}
\caption{ETL 11-inch D784UKFL PMT prototype. In the left image, the discolored area can be seen on the center of the glass
envelope. The right image shows the first dynode of the 12-stage amplifier chain.}
\label{fig:PMTinfo}
\end{figure}
\end{center}

The D784KFLB photomultiplier tube developed by Electron Tube Enterprises Ltd
has an 11" diameter photocathode and twelve linear-focused dynode stages 
(see Figure~\ref{fig:PMTinfo}). It was developed specifically for use in large neutrino detectors
that require high efficiency, fast timing, and low cost. A  single photon
response measurement as well as a relative detection measurement was performed
on a sample of  fifteen prototype tubes. These prototypes have the same photocathode and 
electron optics as the expected final versions, but the glass envelope was manufactured from
Scott 8250 tubing due to the ease in manufacturing of a small number of prototypes.
A known issue with this method and material is that these PMTs have a roughly 25 mm radius doscolored
area  in the center of the front face, where the detection efficiency  of the tube is anticipated
to be impacted. In addition, the glass envelope was not designed for deep submergence
in water. These two issues are not critical in assessing the photon efficiency and 
electron optics, but a correction must be made for the spot when looking at full face
illumination (this is less than a 5\% effect). Note that the final ``fully functional'' 
product (now being manufactured) will be made from water-resistant EU glass and will not have this artifact.

\begin{center}
\end{center}

\section{Single Photoelectron Characterization}

Water Cherenkov and scintillator neutrino detectors require PMTs with excellent timing and charge response. Since the PMTs typically detect no more than a couple photons per event, an accurate characterization of the single photoelectron (SPE) response is critical.

\subsection{Single Photoelectron Characterization Experimental Setup}

\begin{center}
\begin{figure}[h]
\includegraphics[width=0.9\textwidth]{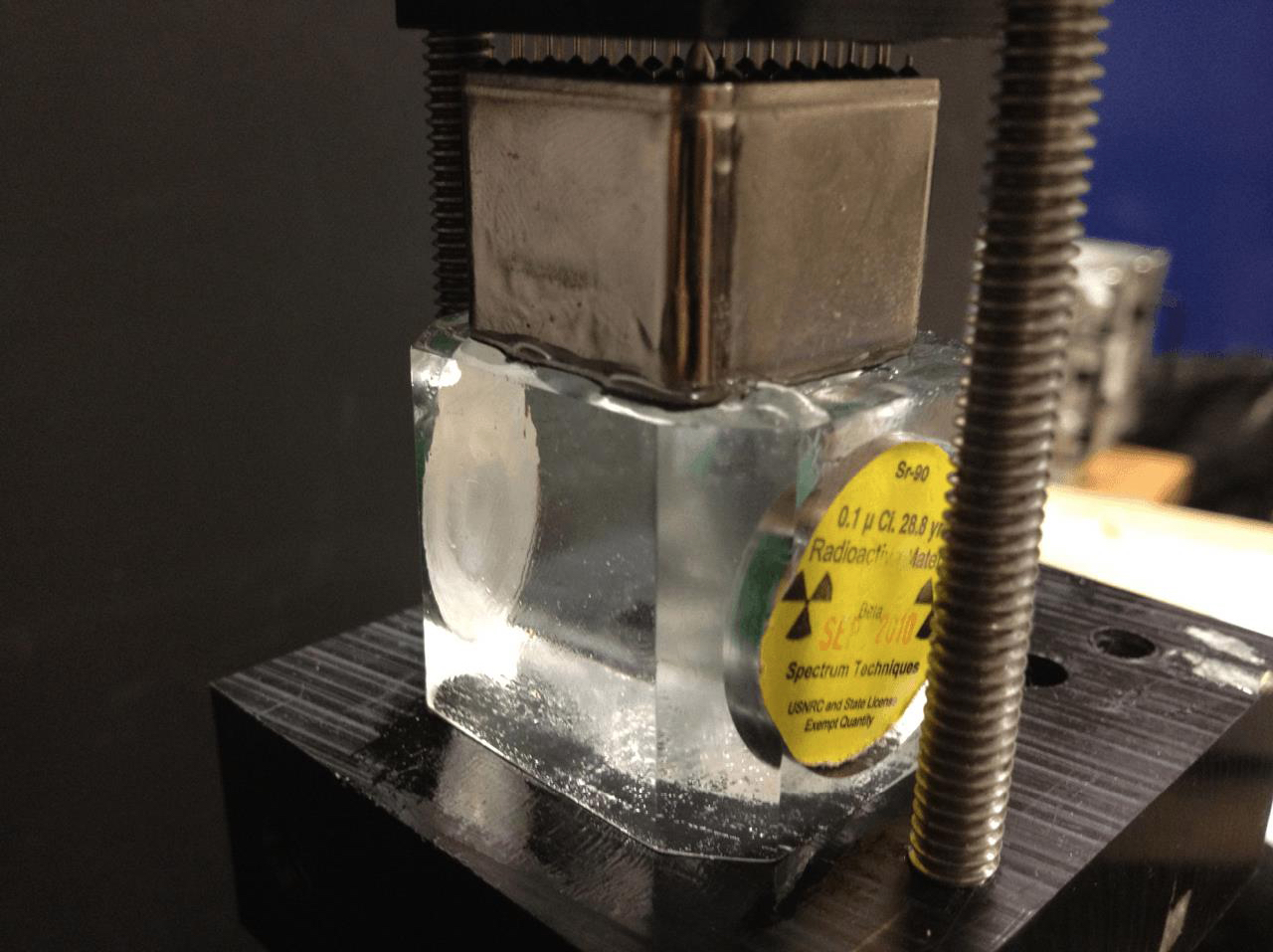}
\caption{The acrylic Cherenkov source. Two $^{90}$Sr disks are embedded into the side of a 1-inch cube, which is optically coupled to a 1-inch R7600-U200 PMT that is used as a trigger.}
\label{fig:CherenkovSource}
\end{figure}
\end{center}

Our measurement of the SPE response relies on a Cherenkov light source, constructed by embedding two 0.1$\mu$Ci discs of $^{90}$Sr in a piece of UV-transparent acrylic machined to an approximately 1-inch cube (see Figure~\ref{fig:CherenkovSource}). The $^{90}$Sr undergoes beta decay to $^{90}$Y with a decay energy of 0.546 MeV and half-life of 28.8 years. The $^{90}$Y also beta decays with a decay energy of 2.28 MeV to $^{90}$Zr, which is a stable isotope. The $^{90}$Y beta travels through the acrylic and creates Cherenkov light. A one-inch high quantum efficiency Hamamatsu R7600-U200 PMT is optically coupled to the acrylic and is used as a fast trigger.

\begin{center}
\end{center}

This Cherenkov source has the advantages of a broadband wavelength spectrum, similar to a water Cherenkov detector, and a very narrow timing distribution. The PMTs to be tested are placed approximately 50 cm from the source in order keep the coincidence rate between hits on the trigger and PMTs below 5\%. The coincidence rate is kept low to reduce the rate of multiple photon hits, which would contaminate the single-photoelectron measurement . At this low coincidence rate, multiple photon events are $<$1\% of the sample, so such contamination is negligible.

\begin{center}
\begin{figure}[h]
\includegraphics[width=0.9\textwidth]{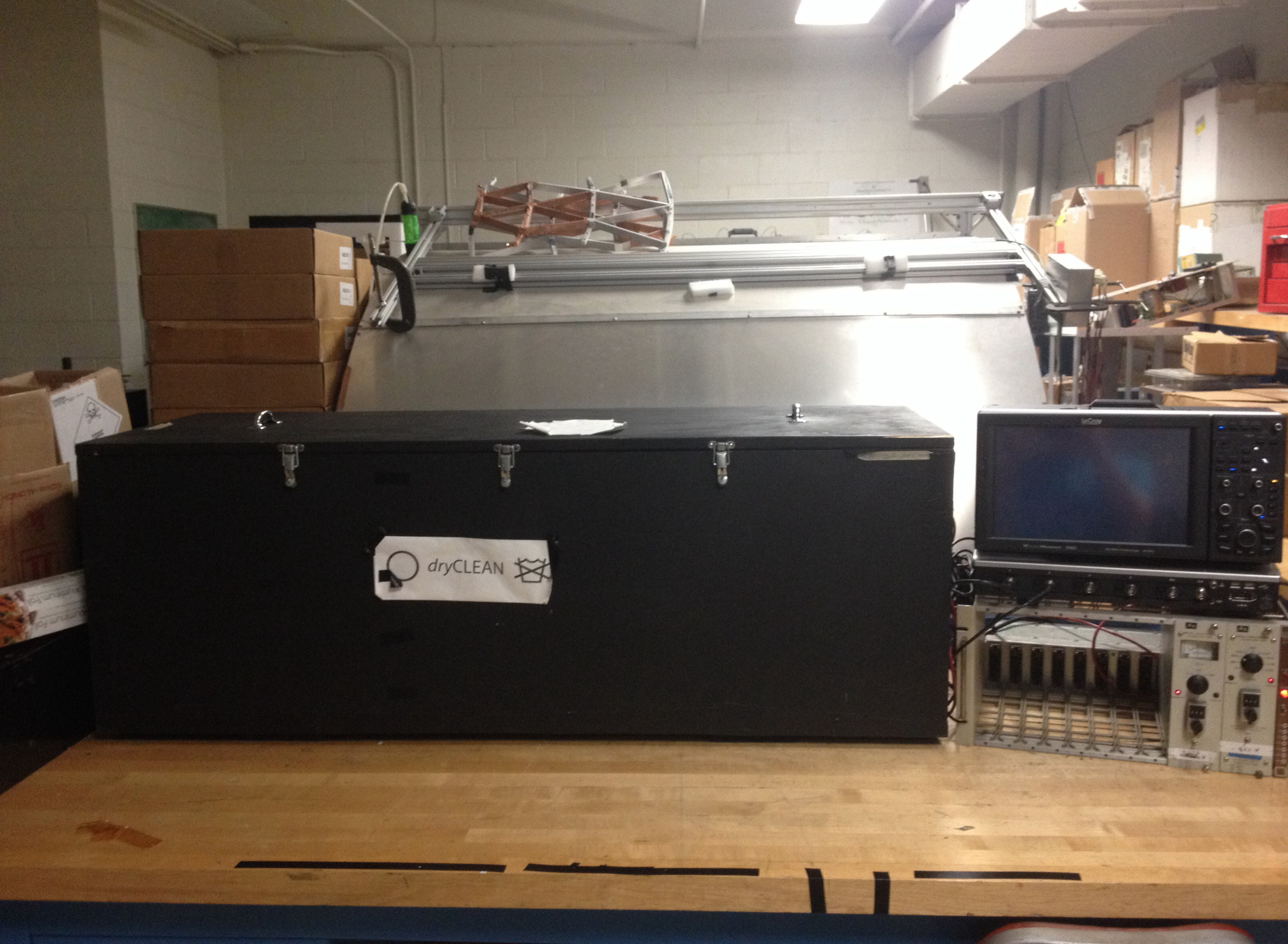}
\caption{The dark box and oscilloscope used in the SPE characterization setup. The dark box is lined with magnetic shielding and covered with felt on both the inside and outside to prevent reflections and light leaks. The oscilloscope is a LeCroy WaveRunner 606Zi 600MHz}
\label{fig:DarkBox}
\end{figure}
\end{center}

\begin{center}
\begin{figure}[h]
\includegraphics[width=0.9\textwidth]{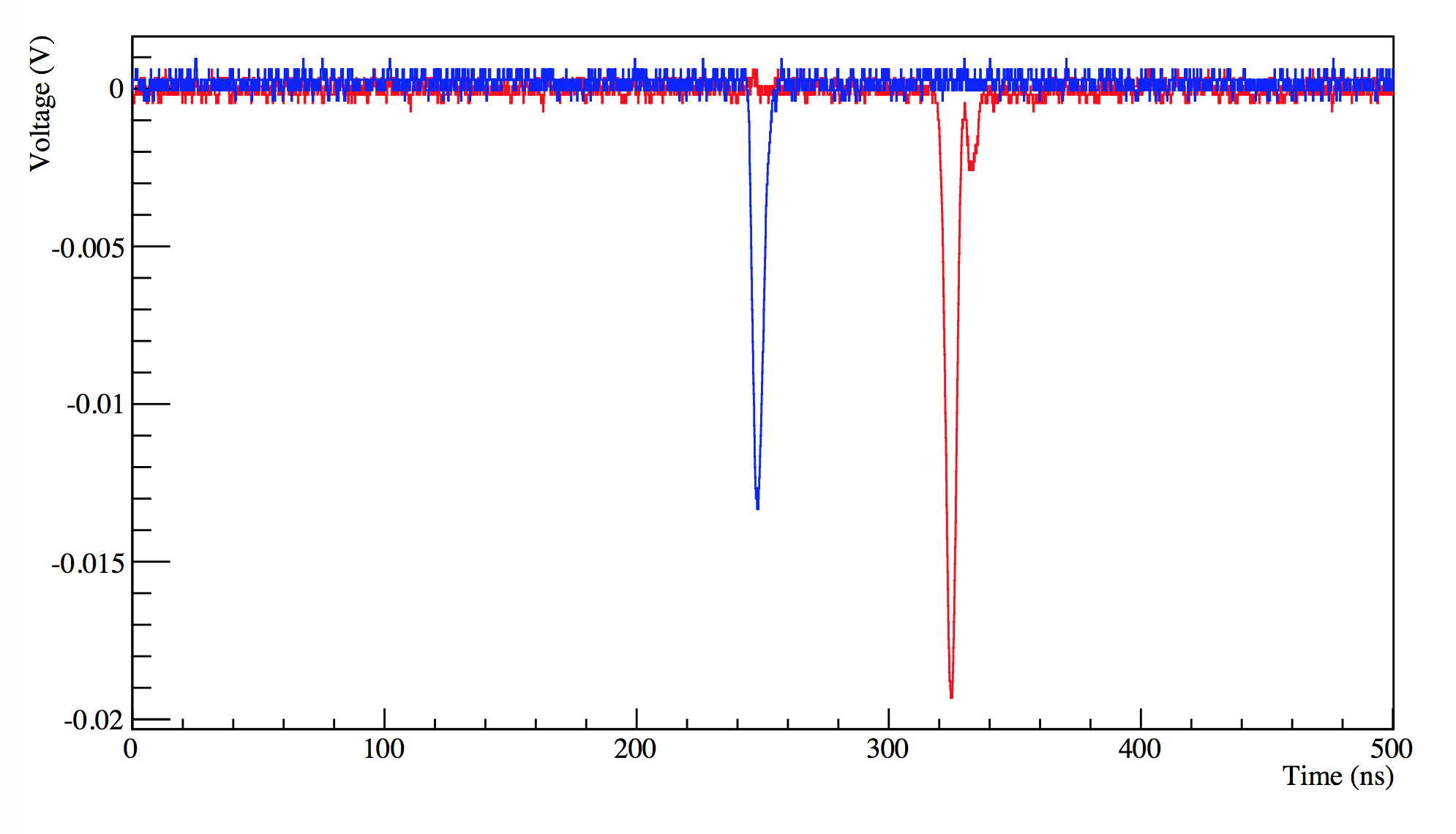}
\caption{The 500 ns digitized waveform is shown in red for the ETEL PMT and that for the fast trigger PMT shown in blue.}
\label{fig:Waveform}
\end{figure}
\end{center}

A dark box shown in Figure~\ref{fig:DarkBox} (size 1.4 x 0.5 x 0.5 m) is used to house the Cherenkov source, trigger PMT, and the PMT being tested. The entire dark box is covered in Finemet magnetic shielding in order to reduce the effects from the EarthÕs magnetic field. A LeCroy WaveRunner 606Zi 600MHz scope was used as the data acquisition system. Waveforms were fetched from the oscilloscope for both the trigger PMT and PMT being tested. Each waveform consisted of 5002 samples at 0.1ns intervals for a total of 500.2ns per trace. In general, data taking consisted of fetching and storing between one and two million triggered events for each PMT. At around a 3\% coincidence rate, 30-60 thousand coincidence events would be stored for each PMT. Thus, the most of the statistical uncertainties on the SPE parameters are below 1\%. Figure~\ref{fig:Waveform} shows a typical waveform for a coincidence event between the trigger and PMT under test.

\begin{center}
\end{center}

\begin{center}
\end{center}

\subsection{Single Photoelectron Characterization Results: Charge}

\begin{center}
\begin{figure}[h]
\includegraphics[width=0.9\textwidth]{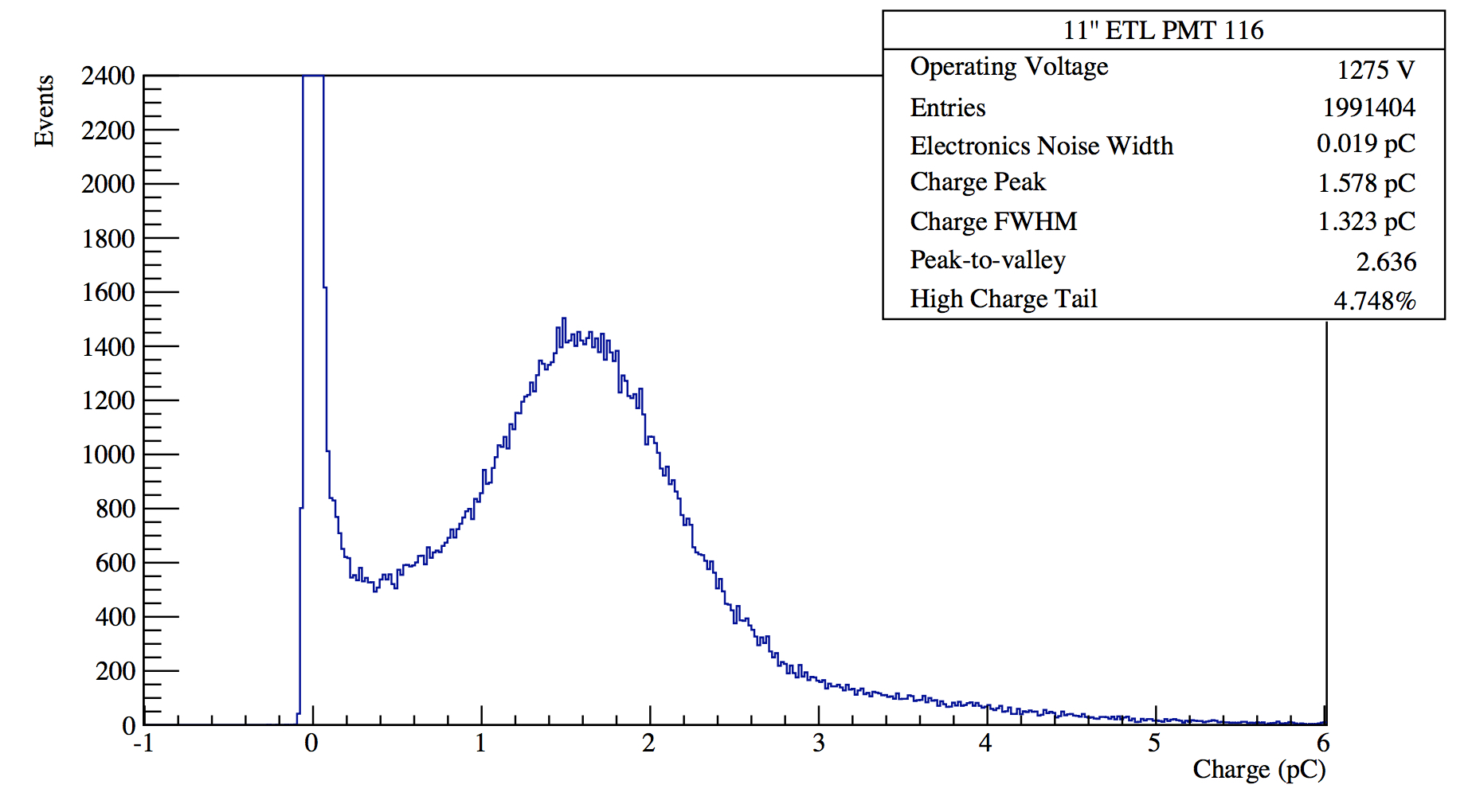}
\caption{The SPE charge spectrum for an 11 inch ETEL PMT. The operating voltage was adjusted to achieve a SPE peak of approximately 1.6pC.}
\label{fig:ChargeSpectrum}
\end{figure}
\end{center}

All stored waveforms were analyzed for fifteen PMTs. The charge spectrum was created by integrating over a fixed 30 ns prompt window corresponding to coincidences on the ETEL PMT. A fixed pre-trigger ÔpedestalÕ window was used to account for the baseline of the trace. Traces with dark hits in the pedestal region were discarded. Based on the dark rate of the PMTs, the cut threw out less than 1\% of traces for each data set. The gains for the PMTs were set to $1\times 10^7$, corresponding to a charge peak at 1.6 pC. Figure~\ref{fig:ChargeSpectrum} shows an example charge spectrum for the PMT serial number 116. This distribution is characterized by a tall narrow ÔpedestalÕ distribution that is created for non-coincidence events by integrating over the electronic noise inherent to the oscilloscope. The SPE charge distribution peaks at about 1.6pC and is characterized by the height of the peak to the depth of the low charge valley. The valley consists of both low charge fluctuations in the SPE response as well as upward fluctuations in the electronics noise. A fit to a Gaussian distribution is used to located the peak of the SPE. Subsequently, a fit to another Gaussian distribution is used to fit the electronics noise. Finally, a fit to a quadratic polynomial is used to find the location of the valley. The following parameters definitions are used:\\

\begin{center}
\end{center}

\noindent {\bf Electronics Noise Width}: A fit to a Gaussian distribution is done around the peak of the pedestal to $\pm N_{half}$ , where $N_{half}$  is the number of bins at the half-height of the peak. The electronics noise width is the standard deviation of this fit.\\

\noindent {\bf Charge Peak}: A fit to a Gaussian distribution is performed in the range 0.67-1.50  Max$_{SPE}$, where Max$_{SPE}$ corresponds to the bin of maximum value in the charge distribution. The charge peak is the charge corresponding to the peak of this fit.\\

\noindent {\bf Charge FWHM}: Using the same Gaussian fit as for the charge peak, the charge FWHM is defined as $2\sqrt{2 \ln{2}}\;\sigma_{SPE}$.\\

\noindent {\bf Peak-to-Valley}: The height of the valley is found using a quadratic fit between $6\sigma$ above the electronics noise width and 0.67 Max$_{SPE}$.
 The height of the charge peak divided by the height of the valley gives the peak-to-valley ratio.\\
 
\noindent {\bf High Charge Tail}: The ratio of the number of events $3\sigma_{SPE}$ above the charge peak to the total number of events above the electronic noise width.\\
 
The results for our measurements are shown in Table 1, which shows the averaged parameters over fifteen tubes. For comparison Tables 2 and 3
 respectively show the results for standard (EQE) and high (HQE) quantum 12-inch Hamamatsu PMTs (R11780). These measurements were done using a similar setup and are described in more detail in \cite{NIM12inch}. 
  To verify that the tests of the ETEL and Hamamatsu tubes was consistent, two HQE R11780 PMTs were tested with exactly the same setup, and the results for the charge spectrum were seen to be completely consistent with \cite{NIM12inch}.\\

\begin{table}
%\label{tab:SPE11}
\begin{center}
\begin{tabular}{ccccc}
\hline
{\bf 11 inch ETEL} & Average & Standard Deviation & Minimum & Maximum \\
\hline\hline
Charge FWHM (pC) & 1.44 & 0.40 & 1.11 & 2.73 \\
Peak/Vally & 2.32 & 0.67 & 1.15 & 3.68 \\
High Charge Tail (\%) & 3.86 & 1.28 & 0.92 & 5.71\\
TTS ($\sigma_{prompt}$) (ns) & 1.98 & 0.17 & 1.79 & 2.47 \\
Late Ratio & 4.51 & 0.74 & 3.0 & 5.76 \\
Operating Voltage & 1330 & 117 & 1183 & 1575 \\
\hline
\end{tabular}
\caption{Summary of SPE characterization results for fifteen ETEL D784KFLB PMTs. All were operated
at a gain of $1\times 10^7$, corresponding to a charge peak of 1.6 pC. }
\end{center}
\end{table}

\begin{table}
\label{tab:SPEEQE}
\begin{center}
\begin{tabular}{ccccc}
\hline
{\bf 12 inch Hamamatsu EQE} & Average & Standard Deviation & Minimum & Maximum \\
\hline\hline
Charge FWHM (pC) & 1.42 & 0.40 & 1.18 & 2.32 \\
Peak/Vally & 2.8 & 0.28 & 2.3 & 3.0 \\
High Charge Tail (\%) & 2.86 & 0.84 & 2.5 & 4.94\\
TTS ($\sigma_{prompt}$) (ns) & 1.37 & 0.15 & 1.20 & 1.6 \\
Late Ratio & 4.48 & 0.32 & 3.93 & 4.92 \\
Operating Voltage & 1848 & 75 & 1740 & 1920 \\
\hline
\end{tabular}
\caption{Summary of SPE characterization results for 7 Hamamatsu R11780 EQE PMTs. All were operated
at a gain of $1\times 10^7$, corresponding to a charge peak of 1.6 pC (from \cite{NIM12inch}).}
\end{center}
\end{table}

\begin{table}
\label{tab:SPEHQE}
\begin{center}
\begin{tabular}{ccccc}
\hline
{\bf 12 inch Hamamatsu HQE} & Average & Standard Deviation & Minimum & Maximum \\
\hline\hline
Charge FWHM (pC) & 1.64 & 0.62 & 1.19 & 3.36 \\
Peak/Vally & 2.24 & 0.27 & 1.78 & 2.76 \\
High Charge Tail (\%) & 3.75 & 0.66 & 2.73 & 5.2\\
TTS ($\sigma_{prompt}$) (ns) & 1.29 & 0.14 & 1.16 & 1.52 \\
Late Ratio & 4.3 & 0.35 & 3.6 & 4.8 \\
Operating Voltage & 1950 & 221 & 1750 & 2500 \\
\hline
\end{tabular}
\caption{Summary of SPE characterization results for 10 Hamamatsu R11780 HQE PMTs. All were operated
at a gain of $1\times 10^7$, corresponding to a charge peak of 1.6 pC (from \cite{NIM12inch}).}
\end{center}
\end{table}

\subsection{Single Photoelectron Characterization: Results: Transit Time Spread}
\label{SPETime}

\begin{center}
\begin{figure}[h]
\includegraphics[width=0.9\textwidth]{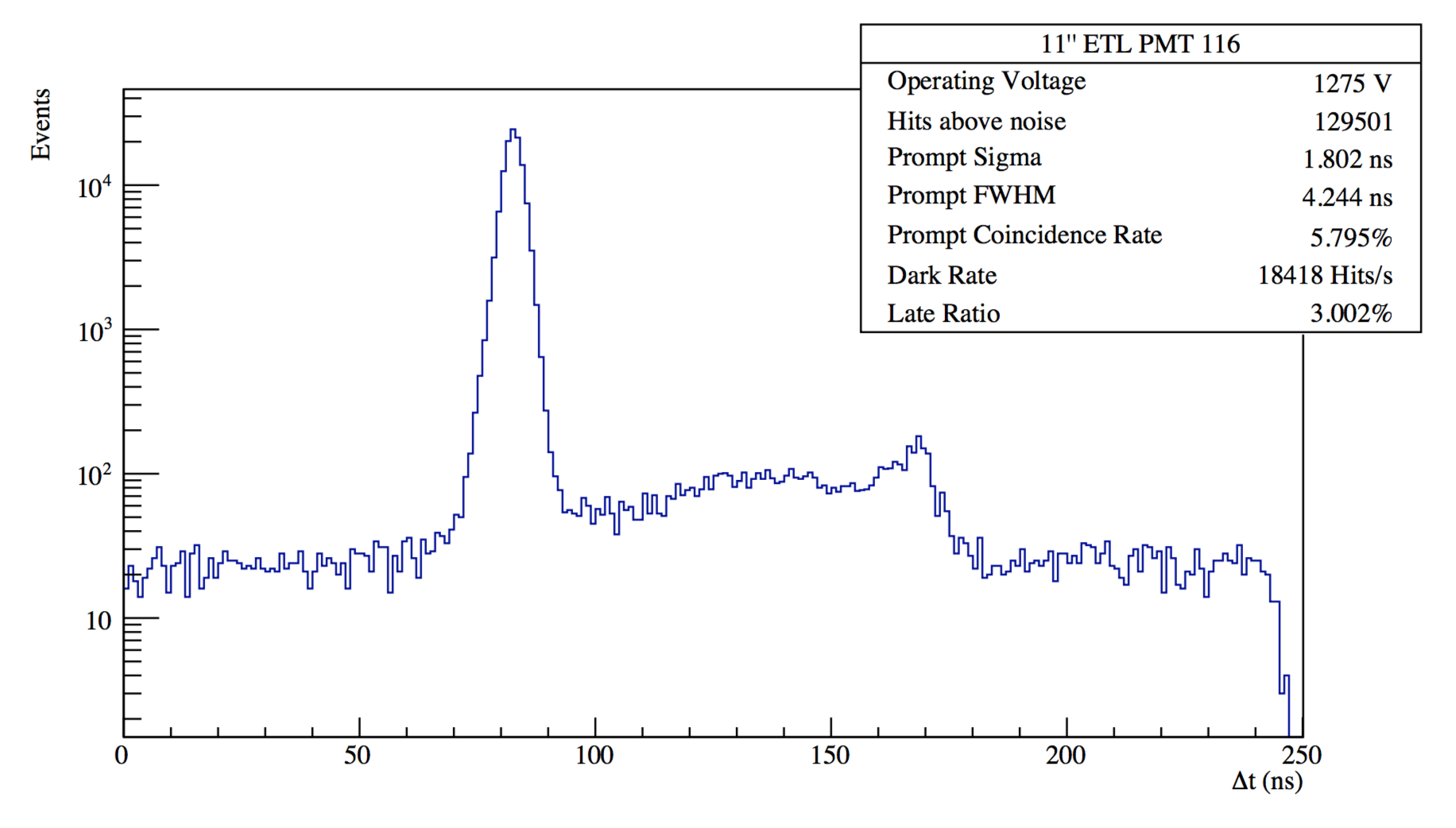}
\caption{The SPE timing profile for an 11inch ETEL PMT.The operating voltage was adjusted for a gain of $1\times 10^7$. Only 30 ns waveform regions with charge above a 0.8pC cut were selected. The timing profile is composed of a prompt peak due to coincidence pulses with the fast trigger, a late peak due to photoelectrons scattering off of the first dynode, and a uniform distribution of dark noise pulses.}
\label{fig:TTS}
\end{figure}
\end{center}

The transit time between the absorption of the photon at the photocathode and the resulting signal at the anode of the PMT varies from photoelectron to photoelectron. Therefore the time of the coincidence pulse relative to the time of the trigger pulse from the fast Cherenkov light source has a multi-component distribution known as the Transit Time Spread (TTS). Since the FWHM of the TTS for the R7600 trigger PMT is only 250 ps,  this adds a negligible jitter to the ETEL PMTs' intrinsic TTS. In order to calculate the TTS for the ETEL tube waveforms with pulses above the noise level were selected by placing a hard cut at 0.8 pC. The waveform of the ETEL PMT was then divided into 30 ns regions, and each region with a charge integral above the noise level was selected for the timing profile. The time of the pulse is defined as the time at which the pulse amplitude
crosses 20\% of its peak height. The transit time is then calculated by subtracting the time of ETEL pulse from the time of the R7600 trigger pulse. The timing profile for 
ETEL PMT number 116 is shown in Figure \ref{fig:TTS}.

The transit time distribution has some clear features.
First is a nearly Gaussian prompt peak corresponding to pulses in coincidence with the Cherenkov source trigger.
Second, a roughly 95 ns long late pulse region, peaking at the end of the distribution. The late pulsing is caused by elastic scattering of 
 photoelectrons off of the first dynode, where they return later to the first dynode roughly two cathode-to-dynode transit times later. Finally, the uniform distribution is
 due to dark noise pulses. The following definitions are used for timing parameters:

\noindent {\bf Prompt Sigma ($\sigma_{prompt}$) and FWHM}: The prompt timing peak is measured by fitting a Gaussian curve to a $\pm 3$ ns region around the maximum bin in the transit time histogram. The prompt FWHM is defined as $2\sqrt{2\ln{2}}\; \sigma_{prompt}$.\\

\noindent {\bf Prompt Coincidence Rate}: The rate of the prompt pulses corrected for accidentals from the dark rate.\\

\noindent {\bf Dark Rate}: The rate of pulses on the PMT that fall outside the late and prompt regions.\\

\noindent {\bf Late Ratio}: The fraction of hits that fall in the late pulse region between $5\sigma_{promt}$ and 95 ns after the prompt transit time peak. Accidentals from dark noise pulses are corrected using this rate.\\

The results for our measurements are shown in Table 1 averaged over fifteen PMTs. Again, these parameters can be compared to the EQE and HQE R11780 PMT shown in Tables 2 and 3 respectively. Most notably, the transit time spread of the ETEL PMTs is on average a factor of 1.5 larger than the R11780. A slightly different analysis was used to generate the ETEL PMT parameters as was used to generate those for the R11780. 

\section{Relative Efficiency Measurements}

A measurement of the absolute detection efficiency of an EQE R11780 PMT was made in \cite{NIM12inch}. Additionally, a relative efficiency
measurement is also described in there. It was found that the detection efficiency of  the HQE R11780 was on average 32\% higher than the EQE PMT. Thus, by directly comparing the ETEL PMTs to the R11780 HQE PMTs we can establish a relative efficiency of the ETEL PMTs to a known standard. This was done by placing the PMTs in the exact same setup and using  the identical Cherenkov source. The coincidence rate, as defined in section \ref{SPETime} for a group of six PMTs is shown in Table 4.
 For the ETEL PMTs a 4-inch diameter double layer of felt was used to cover the entire  discolored part of the front face. Thus, about 13\% of the photocathode area was covered for these measurements. Similar size felt covers were used for all PMTs. By using the felt the comparison is not affected by a possible reduction in the efficiency
due to the discolored area on the ETEL PMTs. 

The coincidence rates were additionally scaled to account for the different size photocathodes of each PMT. As Table \ref{CoinRate} shows, the coincidence rates for the ETEL PMTs are consistent with the rates for the HQE Hamamatsu PMTs. Additionally, the efficiency of the SNO Hamamatsu R1408 PMT is consistent with the knowledge that these PMTs have much lower efficiency than the HQE PMTs. The conclusion is that (at least for this small sample) the efficiency of the ETEL and Hamamatsu HQE PMTs are similar
per square centimeter of photocathode.

\begin{table}
\label{CoinRate}
\begin{center}
\begin{tabular}{c c}
\hline
{\bf PMT Type} & {\bf Coincidence Rate} \\
\hline\hline
12 inch Hamamatsu HQE & 3.42\% \\
10 inch Hamamatsu HQE & 3.46\% \\
11 inch ETEL & 3.01\% \\
11 inch ETEL & 3.31\% \\
11 inch ETEL & 3.63\% \\
8 inch Hamamatsu R1408 & 1.83\% \\
\hline
\end{tabular}
\caption{Corrected coincidence rates for six PMTs.}
\end{center}
\end{table}

\section{Effects Due to Magnetic Fields}

Since photomultiplier tubes rely on shaped electric fields to focus electrons from the photocathode to the first dynode,
they have a well-known sensitivity to magnetic fields even as slight as that of the Earth's. By far, the greatest sensitivity
is to transverse fields (perpendicular to the dynode chain axis), as the Lorentz force will then have the largest magnitude 
and in a direction to move the election off the desired path. This will lower the average collection efficiency when integrated over the 
front face of the PMT. To compensate, some detectors use large scale Helmholtz coils (e.g. Super-Kamiokande~\cite{SKNIM}) or 
mu-metal shields (e.g. Double Chooz~\cite{DCPRL1}). The size of the compensation needed depends on the sensitivity of the 
PMT being used, but in general is worse the larger the PMT format due to the electron travel distances involved.

\subsection{Effects Due to Magnetic Fields Experimental Setup}

\begin{center}
\begin{figure}[h]
\includegraphics[width=0.667\textwidth]{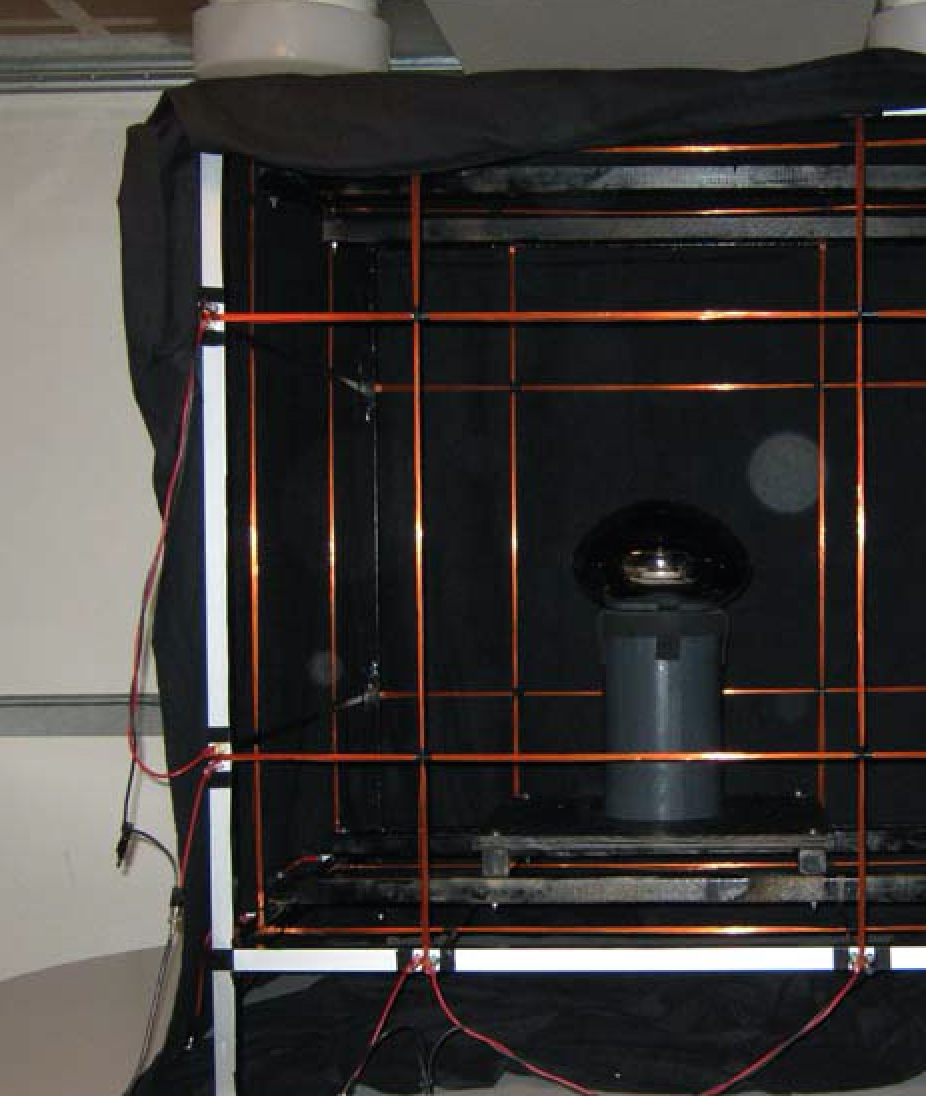}
\caption{The magnetic effects test stand showing tunable triaxial Helmholtz coils set
up inside a large, dark Faraday cage.}
\label{fig:MagLab}
\end{figure}
\end{center}

A test facility with a tunable triaxial magnetic field was used to measure the relative efficiency of two ETEL PMTs. 
This facility is described in detail in \cite{NIM12inch}. Basically, three pairs of one meter wide square copper Helmholtz coils, separated
by 0.5 m, were arranged along three perpendicular axes inside a dark room that is in itself a 
large Faraday cage. Each coil was capable of producing a field of up to 500 mG (similar to the Earth's magnetic field).
The uniformity of the field was measured to be $\pm10$ mG within a 12 inch cube centered on the PMT quasi-spherical bulb, 
as measured with a small Hall probe. 
Figure \ref{fig:MagLab} shows the magnetic field effects test stand.

PMTs were tested by standing them vertically (as shown in Figure \ref{fig:MagLab} where they would view a fast, green
LED flasher roughly 35 centimeters away from the front face. In this coordinate system (centered on the middle of the chills), $+z$
is in the vertical direction upwards, $+x$ is the horizontal direction of pin 1 on the external leads, and $+y$ is the axis perpendicular to 
the first two in a ride-handed system, PMT gain was set to $1\times 10^7$ by adjusting the operating
voltage, and the brightness of the LED was adjusted to give coincidences between the flasher and PMT at a rate
of less than 10\% of the flasher rate. This ensured that virtually all PMT pulses were single photoelectron.
 Two PMTs (serial 120 and 124) were tested. For each, the magnetic field was varied in from -800 mG
 to +800 mG in 50 mG increments for both the $x$ and $y$ directions. The $z$ direction was checked to confirm that
 fields in that direction had little effect.

\subsection{Effects Due to Magnetic Fields Results}

\begin{center}
\begin{figure}[h]
\includegraphics[width=0.9\textwidth]{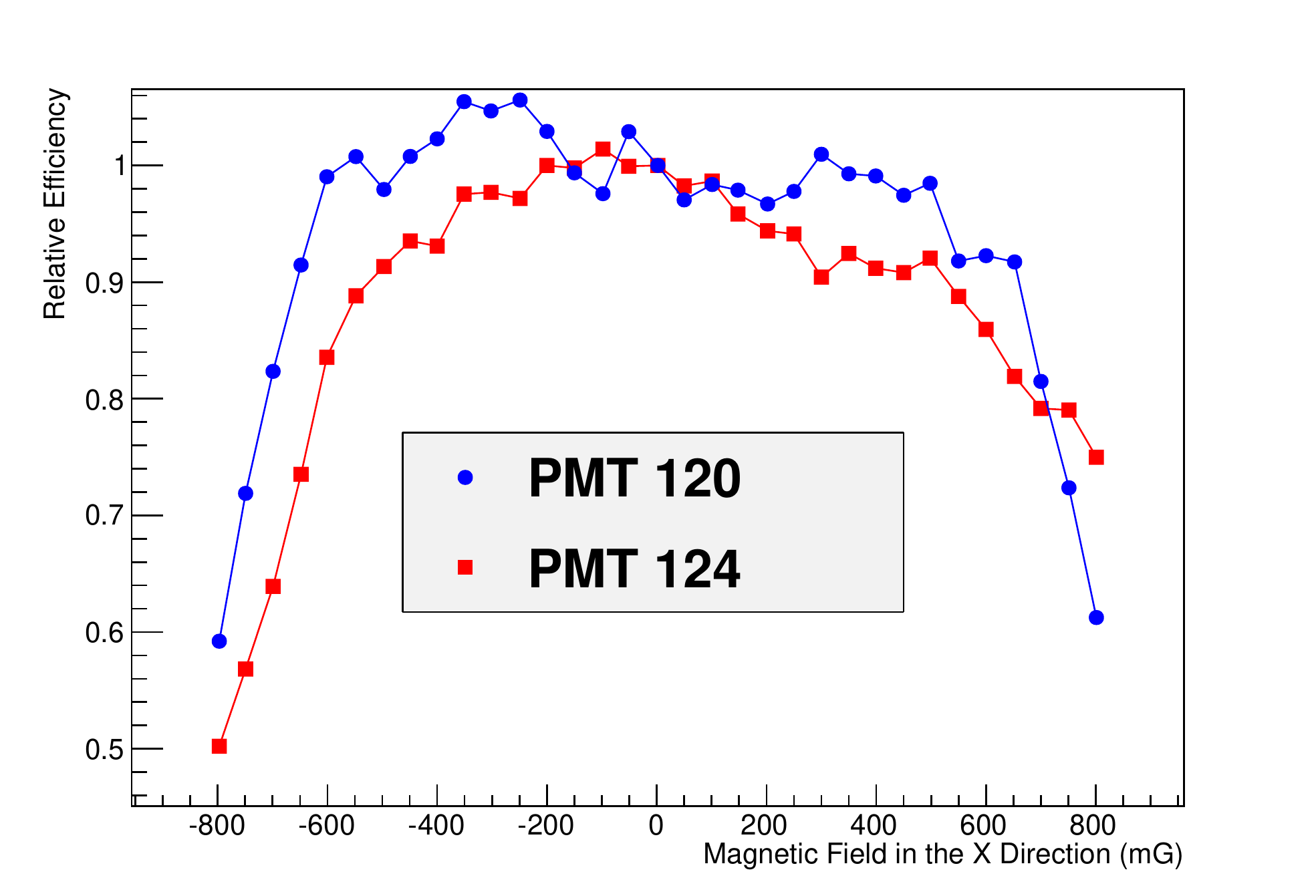}
\caption{Effects of magnetic fields on the relative efficiency in the $x$ direction for
two ETEL 11 inch PMTs.}
\label{fig:MagEffectsX}
\end{figure}
\end{center}

\begin{center}
\begin{figure}[h]
\includegraphics[width=0.9\textwidth]{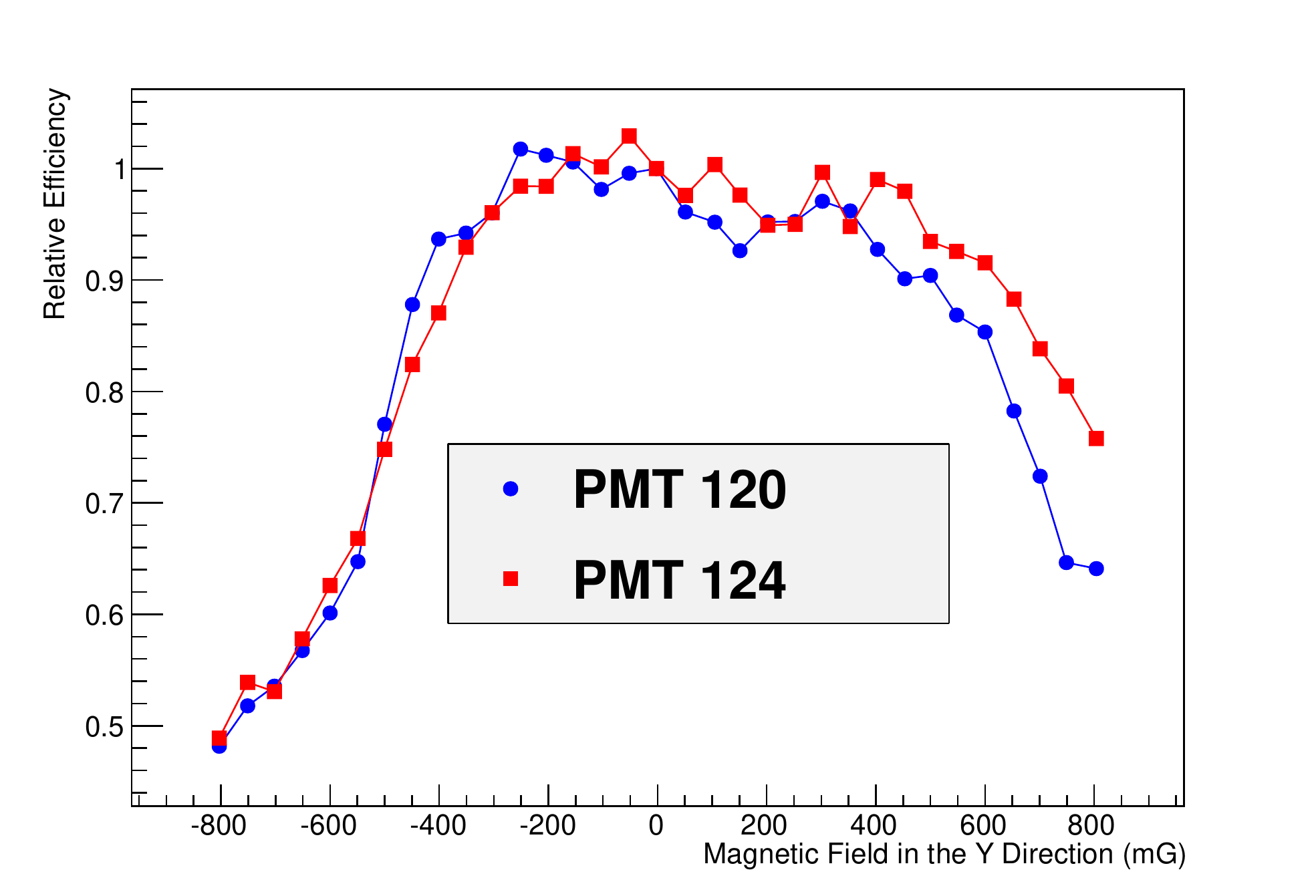}
\caption{Effects of magnetic fields on the relative efficiency in the $y$ direction for
two ETEL 11 inch PMTs.}
\label{fig:MagEffectsY}
\end{figure}
\end{center}

 Figures \ref{fig:MagEffectsX} and \ref{fig:MagEffectsY} show the results in the $x$ and $y$ directions, respectively.
 Both PMTs are similar but not identical at fields similar to that expected from the Earth. Variations could
 be due to differences in the discolored area which was not masked for these tests. It can be seen that
 the ETEL PMTs have less than a 10\% depression in the efficiency for transverse fields of 450 mG, roughly the magnitude of the Earth's field.
  This can be compared to similar tests with the 
 Hamamatsu 12 inch HQE PMT reported in ~\cite{NIM12inch}.
 
\section{Conclusions}

The first set of fifteen D784KFLB 11 inch PMTs have been tested for detection efficiency, timing resolution, dark noise, late pulsing, and susceptibility 
 to transverse magnetic fields. The testing was similar to that done for Hamamatsu EQE and HQE PMTs and reported previously \cite{NIM12inch}. The ETEL
 prototypes have performance comparable to the Hamamatsu PMTs but with slightly large TTS. These first fifteen PMTs were made with Schott 8250 glass tube
 and thus not fully functional in glass transparency or pressure resistance. ETEL is developing fully functional PMTs to be delivered by the end of 2015, and thus
 available for a new generation of large neutrino detectors.  
 
 The authors would like to acknowledge the support received for this work from the National Science Foundation via award 0919550. In addition, students Evan Moore and Jayke Nguyen were supported by the Department of Energy National Nuclear Security Administration under Award Number: DE-NA0000979 through the Nuclear Science and Security Consortium.

This report was prepared as an account of work sponsored by an agency of the United States Government. Neither the United States Government nor any agency thereof, nor any of their employees, makes any warranty, express or limited, or assumes any legal liability or responsibility for the accuracy, completeness, or usefulness of any information, apparatus, product, or process disclosed, or represents that its use would not infringe privately owned rights. Reference herein to any specific commercial product, process, or service by trade name, trademark, manufacturer, or otherwise does not necessarily constitute or imply its endorsement, recommendation, or favoring by the United States Government or any agency thereof. The views and opinions of authors expressed herein do not necessarily state or reflect those of the United States Government or any agency thereof.

\section*{References}

\bibliography{mybibfile}

\end{document}